# Multi-Stream Instability of a Single Long Electron Bunch in a Storage Ring

B.L. Beaudoin, R.A. Kishek, I. Haber, T.W. Koeth, T.M. Antonsen Jr.

Institute for Research in Electronics and Applied Physics (IREAP),
University of Maryland, College Park, MD.

A multi-stream instability is observed experimentally in a longitudinally expanding electron beam in a storage ring. The instability is observed when the beam expands such that its length is several times the circumference of the ring, and portions of the beam overlap. While portions of the beam overlap in physical space, due to the nature of the expansion process, the portions form multiple streams and remain separate in velocity space. The streams become unstable as their number increases and their separation in velocity decreases. An analytical theory predicts the onset of the instability, consistent with simulations and measurements, over a wide range of peak line-charge densities (10.3 pC/m–1.8 nC/m) and bunch lengths. This work extends previous calculations to include the dynamic non-linear elongation of the bunch, with a given initial length, and defines an onset criterion for the filament velocity separation.

Two-stream instabilities in plasmas are ubiquitous, appearing in diverse contexts from inertial fusion plasmas to neutron stars [1-3]. These instabilities are also observed in charged particle beam systems, such as in multiple species accelerators [4-7]. The multi-stream instability, between ion beams and background electron clouds, has been extensively studied, both theoretically [4-5] and experimentally [6]. However, multi-stream instabilities can arise in single specie beams if the beam bunches are allowed to expand under self-fields and spatially overlap, as in the case of multi-bunch injection in rings [7]. In the early 1990s, Hofmann applied a particle-in-cell (PIC) code to simulate the electrostatic coupling that occurs as a multi-bunch beam overlaps, its velocity space mixes, and its phase space filamentary structure disappears [7]. A recent experimental observation of the multi-stream instability at the GSI Helmholtz Centre for Heavy Ion Research (GSI) revealed turbulent current spectra that were believed to be caused by the multi-stream instability of overlapping multiple bunches [8-9].

  This paper extends previous work [7-9] by examining the dynamic non-linear elongation of a single bunch with a given initial length, predicting the time at which the overlapping beam segments become unstable, and determining the range of parameters for which instability appears. Formulas are derived for the lab frame propagation distance and number of filaments at the onset, for a bunch with an initial length and negligible momentum spread. We test the new theory experimentally, over a broad parameter range, demonstrating good agreement. PIC simulations are used to relate the discrepancies between theory and experiment to known particle losses in the latter. This new theoretical framework provides a practical way of determining how long it will take for the instability to develop, if a rectangular beam is left to de-bunch because of space charge.

The experiments presented here expand the parameter range of the GSI experiments [8-9], with bunch lengths reaching up to 41 times longer than the pipe diameter and 300 times longer than the beam radius at its minimum. Although this work focuses on a single expanding bunch, the physics can be extended to multi-bunch injection [7, 10] or to longitudinal stacking of short rectangular bunches in a linac.

Experiments were conducted using the University of Maryland Electron Ring (UMER). Table I describes the measured and calculated parameters over which the data were taken. An aperture wheel, downstream of the anode was used to inject different bunch currents into the ring. The onset of instability was observed employing a resistive wall current monitor, installed 6.4 m around the ring from the injection point, to measure the current versus time for the different beam currents and initial beam pulse durations. A sample trace is shown in Fig. 1. The beam travel time around the ring is roughly 0.2 $\mu s$, the initial beam pulse duration is a fraction of this time, so the measured current initially oscillates at 5 MHz. As the beam expands longitudinally and overlaps itself the oscillation amplitude decreases and the beam current becomes steady. At approximately 17 $\mu s$ when it is estimated the beam has overlapped itself 2-3 times the instability appears abruptly.

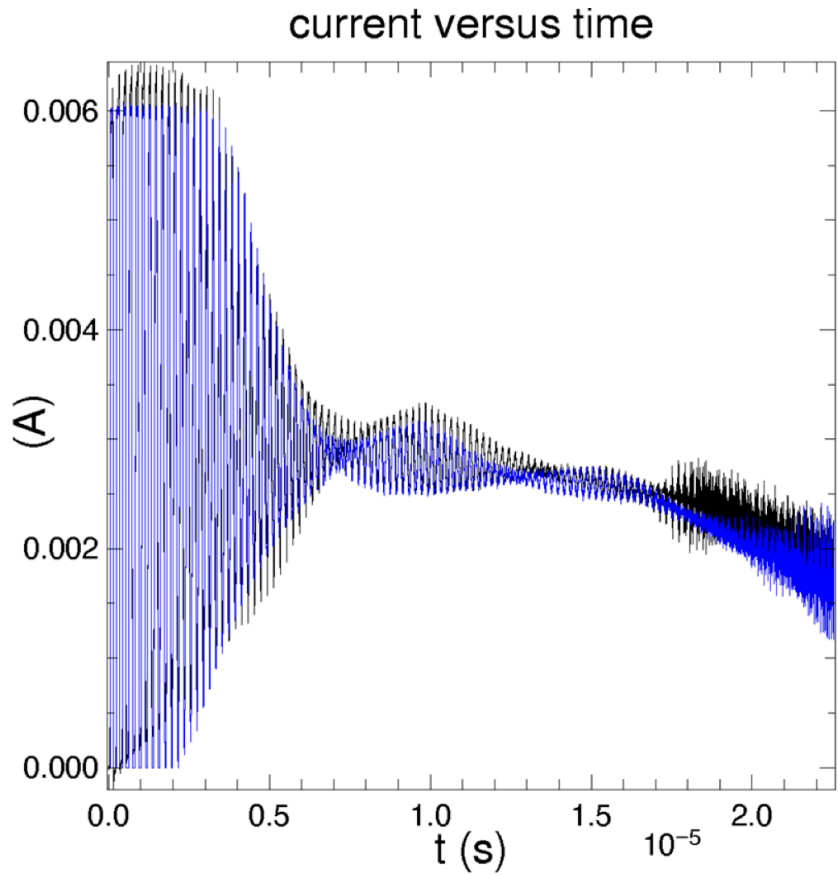

Figure 1 Measured wall current versus time. Initially the beam is in a tight bunch and the wall current oscillates at 5 MHz, the reciprocal of the travel time around the ring. As the beam expands and overlaps itself the oscillation amplitude diminishes. At about 17 $\mu s$ the instability appears.

The instability was also simulated using the PIC code WARP [19]. Simulations of the 10 keV beam were performed in the beam frame by "straightening the beam" and using periodic boundary conditions in the direction of propagation, with a period equal to the ring circumference. As a further approximation, the alternating-gradient focusing lattice was replaced with an average and constant inward linear focusing field. This assumption permitted the use of a two-dimensional RZ field-solver with the simulated beam axisymmetric around the pipe center. The number of cells in r and z was 64 and 256, and the total number of macro-particles in the simulation was 10 million with a step of 10 cm or approximately 1.71 ns. The simulation also assumed an initial beam with a rectangular current distribution and constant velocity in a conducting pipe with a longitudinal thermal momentum spread of Δp/p = 0.00257. We incorporated the measured charge loss rate applied uniformly in r and z to account for beam scrape-off due to 3D effects. It has previously been shown that these assumptions can be used to accurately reproduce details of the longitudinal dynamics [17-18].

The sequence of longitudinal z-$v_z$ phase space plots (shown in [Fig. 2]), illustrates the expansion and overlapping of the beam, and the growth of the multi stream instability.

The longitudinal dynamics of these high-intensity electron beams exhibit many characteristics of bounded non-neutral plasmas including perturbations associated with space-charge waves [11]. The space charge waves are essentially one-dimensional plasma waves for which the effective plasma frequency is wave number dependent due to the presence of the conducting tunnel surrounding the beam. The effect of the tunnel is to shield the interaction between portions of the beam that are axially separated by more than the diameter of the tunnel. The result for perturbations that vary on an axial scale greater than the tunnel diameter is $E_z(z) = -\partial \Phi / \partial z$, where the effective potential, which is depressed from that of the wall, depends on the local line charge density, $\lambda_0 n(z)$, where n(z) is a dimensionless density profile

$$\Phi(z) = g\lambda_0 n(z)/(4\pi\varepsilon_0 \gamma_0^2). \qquad (1)$$

Here g is a dimensionless geometric factor that depends on the beam radius and the wall radius [12-13]; $g = 2\ln(r_w/r_B)$ where $r_w$ is the wall radius and $r_B$ is the beam radius

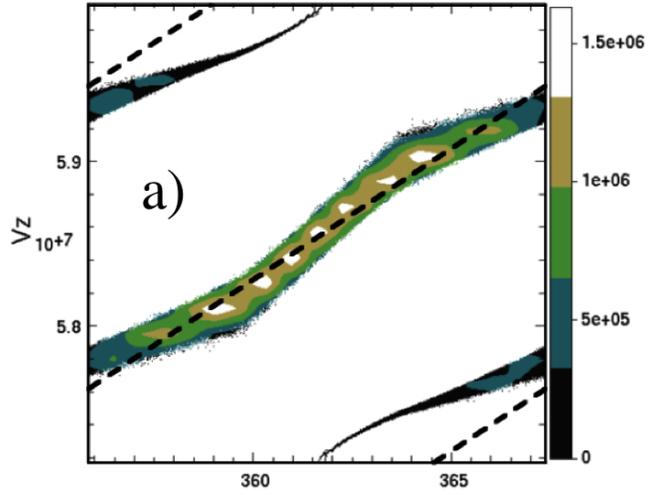

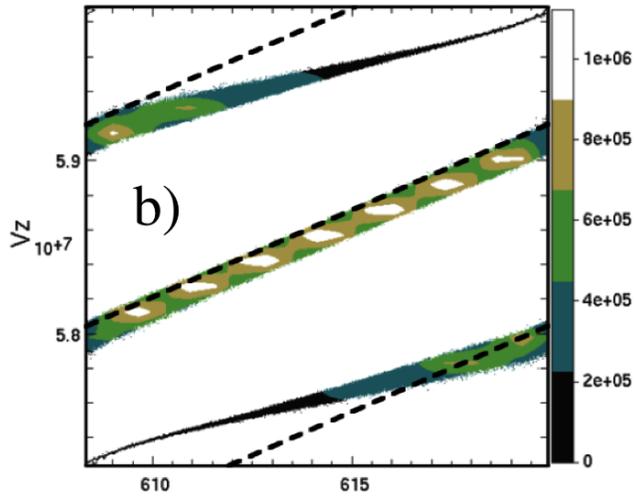

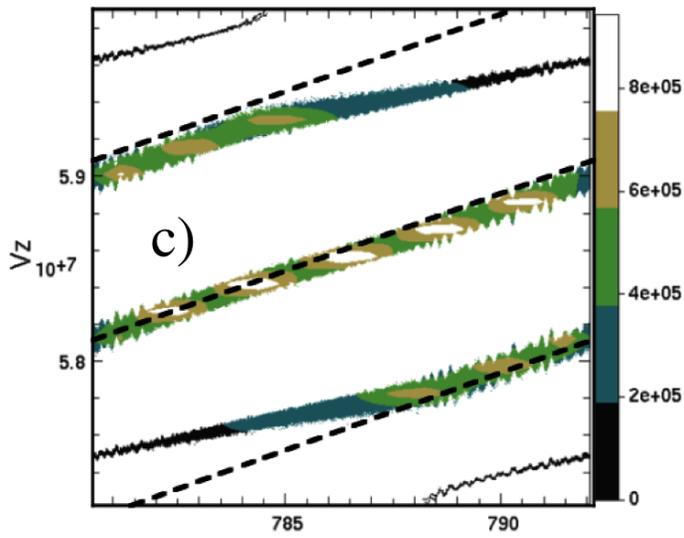

**Figure 2.** Axial velocity versus axial distance from simulations of beam dynamics. Plots are shown for three different propagation distances a) 361.6m b) 614.1 m and c) 786.3 m. The dashed lines show the velocity profile chosen for evaluation of the dispersion relation. The instability in the simulation can be seen in the axial modulation of the beam in frame c)

and $r_w/r_B \gg 1$ as is the case in UMER. The quantity $\gamma_o$ is the Lorentz factor that has been added to account for the azimuthal magnetic field perturbation, and $\varepsilon_o$ is the permittivity of free space. When the interaction potential (1) is inserted in the 1D, nonrelativistic cold fluid equations for a beam of electrons with velocity v, charge $q$, mass $m$, and line charge density $\lambda_0 n$, the resulting system consists of the continuity equation, $\partial n/\partial t + \partial nv/\partial z = 0$, and the force balance equation $\partial v/\partial t + v\partial v/\partial z = -c_s^2 \partial n/\partial z$, where the 'sound speed' is given by

$$c_s^2 = \frac{q\lambda_0 g}{4\pi\varepsilon_0 m\gamma_0^5}. \qquad (2)$$

The dispersion relation for perturbations with frequency $\omega$ and axial wavenumber $k$ of a one-dimensional beam with equilibrium velocity $v_0$ is $\omega = kv_0 \pm kc_s$.

The multi-stream instability reported in this paper, results from the longitudinal expansion of the initial rectangular bunch from the residual $E_z$ fields at the beam-ends [14-15]. Assuming that the initial bunch has a constant line-charge density and velocity, $v_0$, the beam-ends initially elongate at a speed of approximately $2c_s$ [14] relative to the average velocity while at the same time a rarefaction wave propagates back towards the beam center at speed $c_s$. This can be seen in frame a) of Fig. 2. The rarefaction disturbance reflects at the center of the beam and the density profile becomes diffuse as the beam expands and overlaps as illustrated in frames b)-c) of Fig. 2. A key feature of this time evolving system is that as the beam overlaps itself the individual streams occupying the same location become closer and closer to each other in velocity. Instability appears only when the velocity difference becomes small enough. In the case of Fig. 2 the instability can be seen in frame c).

Using these images, we model the long-time dependence of the axial profiles of the beam density and velocity by a similarity solution. We take for the normalized density, $n(z,t) = (L_i / L(t)) \hat{n}(\xi)$, and for the velocity, $v(z,t) = \bar{v} + \Delta v \xi$, with $\xi = 2(z - \bar{v}t)/L(t)$ where $|\xi| \leq 1$ measures distance from the center of the beam, and $L(t) = L_i + 2\Delta v t$ is the length of the expanding beam whose initial length is $L_i$. We note that for this similarity solution we have chosen velocity to be a linear function of distance with a time evolving slope. We see from the results of the simulations shown in Fig. 2 that initially (Fig. 2a) velocity is piece-wise linear with two slopes, but then later (Figs. 2b and 2c) the velocity becomes more linear with a single slope. The beginning of the instability growth can be seen in Fig. 2c.

To relate the density and velocity profiles we consider the cold fluid continuity and momentum equations. They conserve particle number and energy, $N = \int dz\, n(z,t)$ and $U = (1/2) \int dz \{nv^2 + c_s^2 n^2\}$. Initially, the normalized density is unity over a length $L_i$, the velocity is zero in the beam frame, and thus, $U = L_i c_s^2 / 2$. Later, when the similarity solution applies the energy is dominated by the kinetic energy, $U = L(\Delta v/2)^2 \int_{-1}^{1} d\xi \hat{n}(\xi) \xi^2$. From the line-outs of the phase space density of Fig. 2 we model $\hat{n}(\xi) = (15/8)(1 - 2\xi^2 + \xi^4)$, with the coefficient chosen to keep the integrated density equal to the initial density. Figure 3 shows a plot of the model density $\hat{n}(\xi)/\hat{n}(0)$ versus normalized axial distance $\xi$ compared with line-out data taken from the simulations producing Fig. 2b.

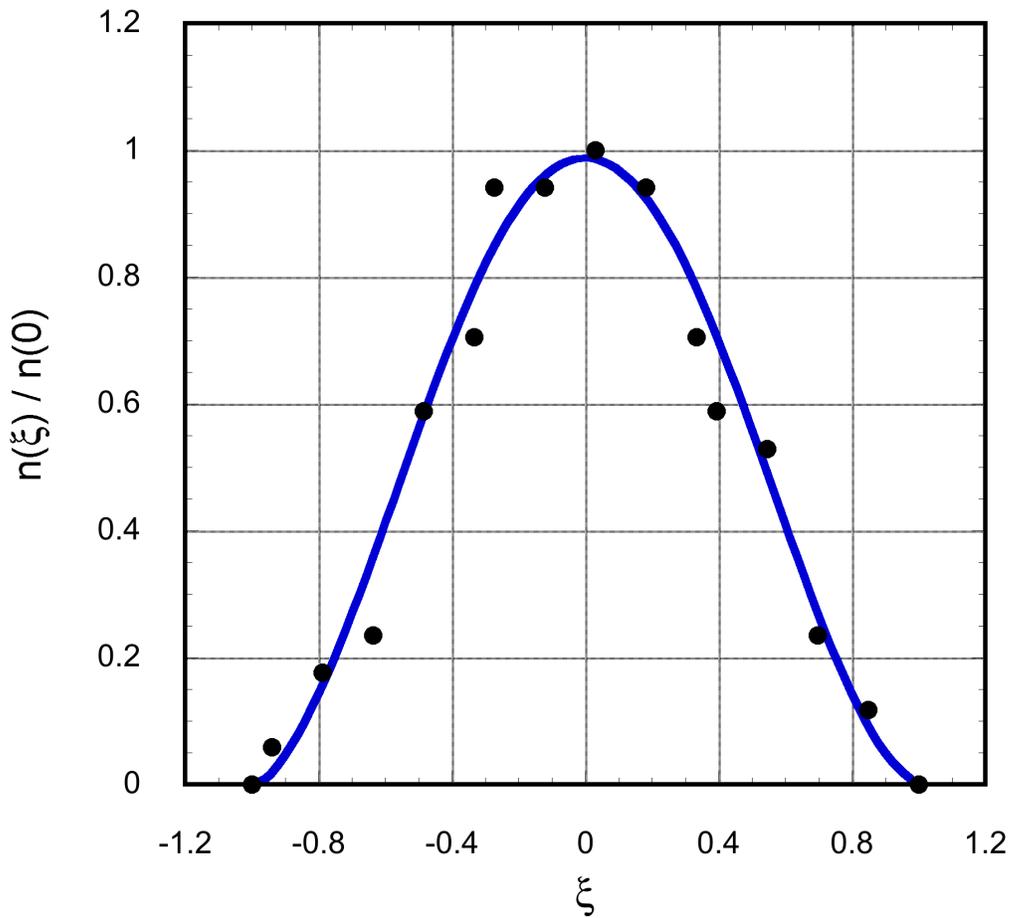

**Figure 3**. Plots of normalized density versus normalized distance. The points are extracted form line-outs of simulation results shown in Fig. 2b. The solid line is the fitted profile.

Equating the initial and final energies then gives $\Delta v = \sqrt{7} c_s \simeq 2.6 c_s$. The implied velocity profile is then superimposed as dashed lines on frames b) and c) of Fig. 2. We will use this in our modeling of wave growth.

We now formulate a dispersion relation for beam plasma modes in a multistream beam. As the length of the beam increases beyond the circumference of the ring, $L_C$, portions of the beam overlap. If we consider a slice of the beam located at $\zeta = 2(z - \bar{v}t)/L_C$ with $|\zeta| \leq 1$, it will coincide with other slices of the beam located in our self-similar coordinates at $\xi_n = (\zeta + 2n)L_C / L(t)$ for all integers $n$ giving $|\xi_n| \leq 1$. These slices have velocities $v_n = \bar{v} + \xi_n \Delta v$ and normalized densities $\hat{n}_n = \hat{n}(\xi_n)$. Thus, a generalization of the beam dispersion relation accounting for the multiple streams with multiple velocities and densities takes the form,

$$1 = \frac{k^2 c_{s0}^2}{1 + k^2 r_w^2} \frac{L_i}{L} \sum_n \frac{\hat{n}_n}{(\omega - k v_n)^2} = \frac{L_i / L}{1 + k^2 r_w^2} \sum_n \frac{\hat{n}_n}{(\Omega - v_n / c_{s0})^2}, \quad (3)$$

where the sum is over the beam slices at the location defined by $\zeta$, and $\Omega = \omega/(k c_{s0})$ is a normalized frequency. Except for the term $1 + k^2 r_w^2$ in the denominator, the dispersion relation follows directly from the cold fluid equations and the interaction potential defined by Eq. (1) and Eq. (2). The term in the denominator has been added and accounts for the fact that the potential defined by (1) and (2) only applies to perturbations with axial wavelength much greater than the pipe radius $r_w$. If this term is left out, the solutions of the dispersion relation for the normalized frequency $\Omega$ are independent of wavenumber, meaning the actual frequency and growth rate are proportional to wavenumber, and consequently the model predicts arbitrarily large growth rate as soon as instability occurs. As will be seen, the large wavenumber correction, accounting for the nonzero tube radius, limits the maximum growth rate. Another limiting factor is thermal spread. We may account for thermal spread in the beamlets by subtracting $i|k|v_{th}$ from the frequency.

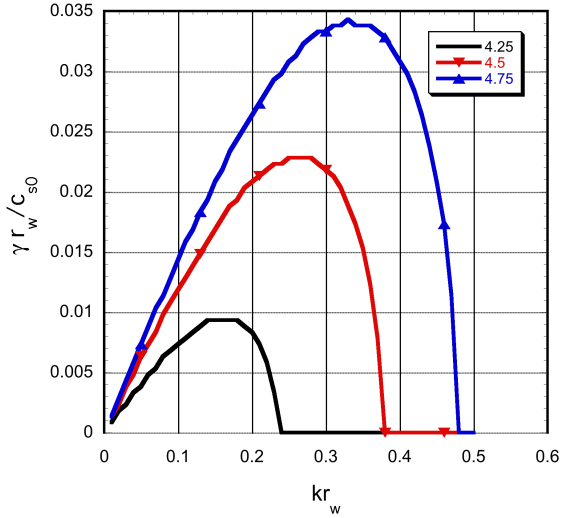
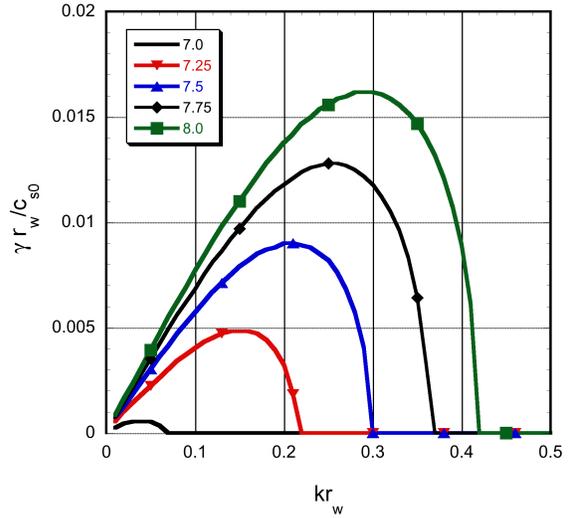

Figure 4a. Normalized growth rate versus normalized wave number at three tines during the simulation. Here $\eta = L_i / L_c = 0.5$. The three curves correspond to cases, in which, the beam has expanded to the indicated fraction of $L_c$.

Figure 4b. Normalized growth rate versus normalized wave number at three tines during the simulation. Here $\eta = L_i / L_c = 0.25$. The five curves correspond to cases, in which, the beam has expanded to the indicated fraction of $L_c$.

Here we have assumed that the velocity distribution for each portion of the beam is Lorenzian, with a width $v_{th}$. However, as seen in Fig 2 the beam is relatively cold in that the width of the velocity distribution for individual streams is much smaller than separation in velocity between streams.

Dispersion relation (3) has several basic independent parameters. These are the normalized wave number, $kr_w$, the length of the beam as measured in circumferences, $L/L_c$, the normalized initial beam length $L_i/L$, and the location of the beam slices that interact as measured by $\zeta = 2(z - \bar{v}t)/L_C$. Instability will not occur before the beam expands to $L > L_c$ and streams overlap. When streams first overlap they are separated in velocity by 5.4 $c_{s0}$, which is too large a separation for instability to appear. However, as the beam continues to spread the separation in velocity between streams decreases (see Fig. 2) and instability can appear.

Figures 4a and 4b show normalized growth rate plotted versus normalized wavenumber at times for which the beam has expanded to different lengths ($L/L_c$). Figure 4a corresponds to the case in which $L_i/L_c = 0.50$ and Fig. 4b corresponds to $L_i/L_c = 0.25$. The general behavior of the growth rate is the same in both cases. Once the beam has expanded by a sufficient amount, growth first appears at long wavelengths. Then as the beam expands further the range of unstable wavenumbers increases as does the growth rate at a fixed wavenumber. Instability occurs earlier for beams that initially occupy a larger fraction of the circumference than those that don't. In the case of Fig. 4a ($L_i/L_c = 0.5$), instability appears when $L_i/L_c \simeq 4$, and in the case of Fig. 4a ($L_i/L_c = 0.25$), instability appears when $L_i/L_c \simeq 7$. The wavelength of perturbations that is expected to be observed will depend on the level of noise from which the perturbations grow. For very low noise levels the beam will have a longer time to expand before the perturbations have grown and saturated than it will for high noise levels. As a consequence shorter wavelengths can be expected in the low noise case. The growth rates in Figs. 3 are normalized to the sound speed divided be the wall radius. The time for the beam to increase by one circumference in length is roughly the circumference divided by the sound speed. Given that the circumference is three orders of magnitude greater than the tube radius one can expect that perturbations will be observable shortly after instability first occurs.

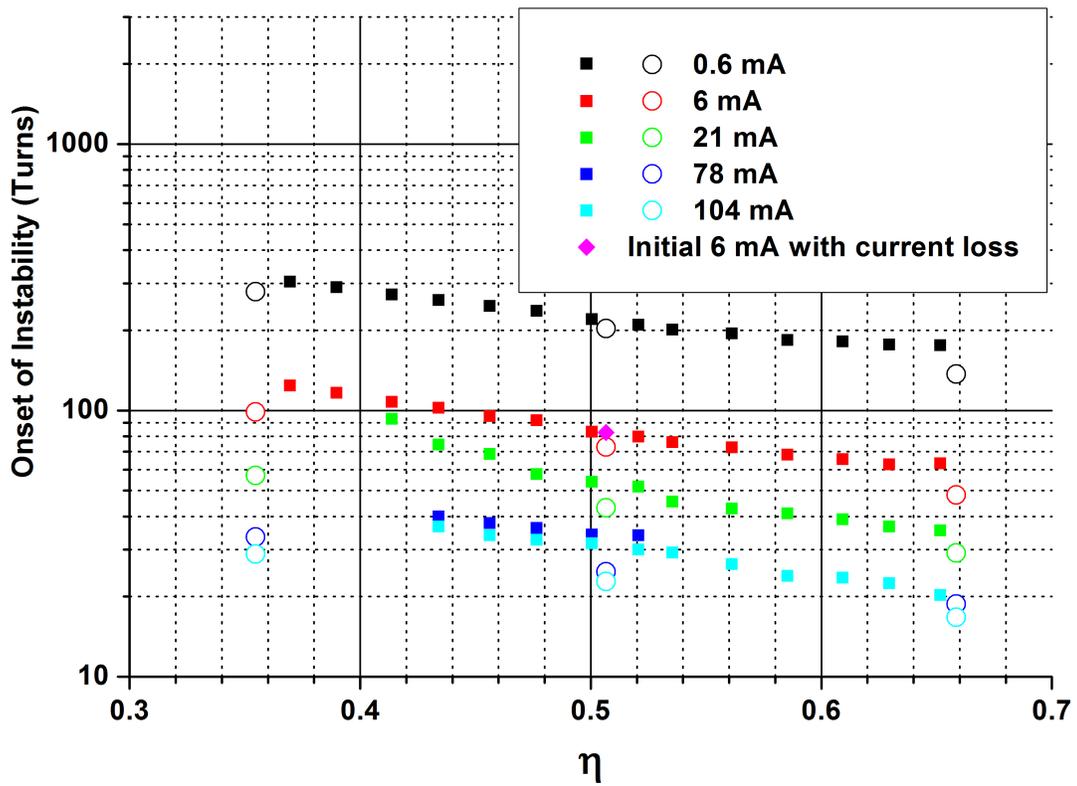

**Figure 5** Experimental measurements (shown as solid squares) of the onset, plotted as a function of $\eta$, the fill factor or ratio of pulse length to machine circumference. Measurements are compared with theoretical calculations (continuous curves) and WARP simulations (circles) for five different beam line-charge densities.

We now compare the measured, simulated, and predicted times for the appearance of the instability as a function of beam parameters. The measured onset was estimated for different parameters from plots of the type shown in Fig. 1. The results of a series of measurements and comparison simulations are shown in Fig. 5.

Here we plot the number of turns (essentially the time) until the onset of instability versus the normalized initial bunch length $\eta = L_i / L_c$ for several different initial beam currents. The experimental times are obtained from plots of the type shown in Fig. 1, and are plotted as solid symbols. The time to instability from the simulations are plotted as open symbols. The simulations and experiments show the same dependence on beam current and initial bunch length.

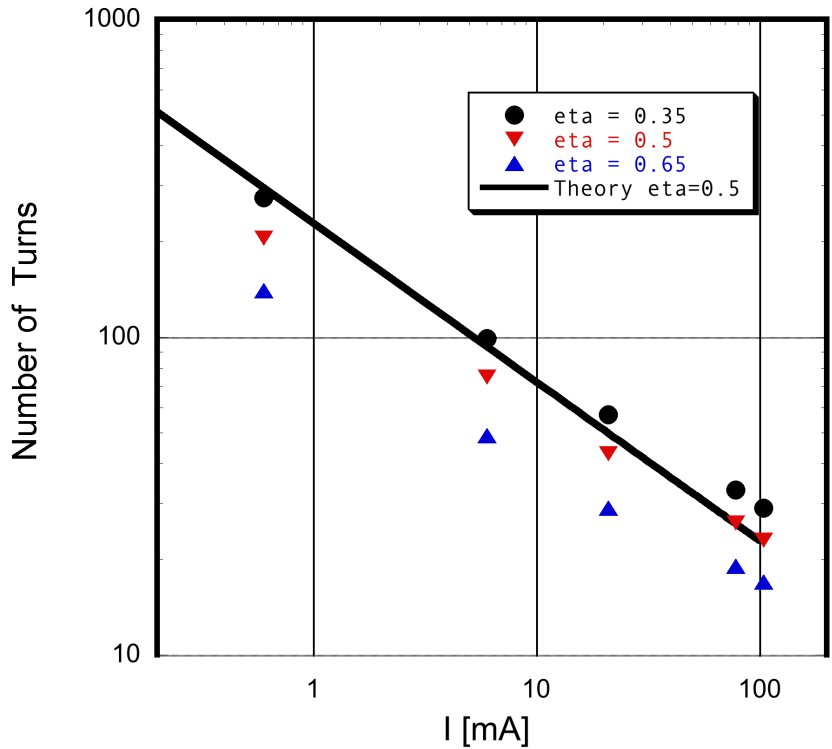

**Figure 6** shows plots of time to instability versus beam current for different values of initial bunch length. The simulation data are the same as in Fig. 5 and are shown as symbols. The theory, based on the scaling implied by Eq. (3) and the results of Fig. 4a is plotted is a continuous curve. Both types of curve show the same scaling.

Previous measurements had resolved a charge loss mechanism over multiple turns that contributed to the slowing of the erosion rate and thus the wrapping of the beam ends [17-18]. This slowing of the longitudinal wave velocity has a direct impact on the instability, as it delays the onset before the filaments are separated by $c_s$, requiring the beam to propagate for a longer distance or more turns in the ring. Incorporating the measured charge loss curves into the PIC simulations (through particle weights adjustments), allowed us to reproduce the loss rate from turn-to-turn. Using this approach, we obtained good agreement for a beam with an initial line-charge density of 103 pC/m and fill factor of 0.506 [Fig. 5]. When charge loss was included, as shown by the magenta diamond, we were able to resolve the simulated delayed onset at 16.3 µs or 82.6 turns. Comparing these results with measurements, for a fill factor of 0.5004, indicate the onset to occur in 83.2 turns. The 0.6 of a turn delay in the measurement value agrees with that fact that a shorter injected bunch requires more time for the onset to occur. The loss curves are also derived experimentally and only estimate the complex three-dimensional dynamics using an RZ geometry. When no charge loss is accounted for, the simulated onset occurs in 72.9 turns, which is an 11.7% decrease from the delayed onset due to charge loss.

To conclude, we have shown experimentally, computationally and analytically, that the onset of the multi-stream instability in a single long electron bunch, corresponds to earlier onsets for bunches with higher total charge. We have derived a simple theory and verified it with PIC simulations, predicting the onset of the instability when the filamented velocity separation is equal to a longitudinal wave velocity. We also extended previous definitions to include the dynamic non-linear elongation of a bunch given an initial length. When including the effects of charge loss from turn-to-turn, we also reproduced in simulation, the delayed experimental onset measurements. The slow loss of charge, effectively delays the longitudinal wrapping of the beam as well as the onset of the instability.

This work was supported by the U.S. Dept. of Energy, Offices of High Energy Physics and Fusion Energy Science, and by the Dept. of Defense, Office of Naval Research and the Joint Technology Office.

# REFERENCES

[1] R.A. Jameson, in Advanced Accelerator Concepts, AIP Conf Proc 279, edited by J.S. Wurtele (AIP, New York, 1993), p.969.

[2] N. Andersson, G.L. Comer, and R. Prix, Phys. Rev. Lett. **90**, 091101 (2003).

[3] N. Anderson, G.L. Comer, and R. Prix, Monthly Notices of the Royal Astronomical Society 354, p.101-110 (2004).

[4] R.C. Davidson and H. Qin, *Physics of Intense Charged Particle Beams in High Energy Accelerators* (Imperial College Press, Covent Garden, London, 2001).

[5] R.C. Davidson, E.A. Startsev, M. Dorf, I.D. Kaganovich and H. Qin, Proc. 2009 Part Accel Conf, Vancouver (IEEE, New York, 2009), p. 1525.

[6] R.J. Macek, *et al.*, Proc. 2001 Part Accel Conf, Chicago (IEEE, New York, 2001), p. 688.

[7] I. Hofmann, Part. Accel., **34**, p.211-220 (1990).

[8] S. Appel, T. Weiland and O. Boine-Frankenheim, Proc. 2010 International Part Accel Conf, Kyoto (IEEE, New York, 2010), p. 1916.

[9] S. Appel and O. Boine-Frankenheim, Physical Review Special Topics – Accelerators & Beams **14**, 054201 (2012).

[10] J. Beebe-Wang, Proc. 1999 Part Accel Conf, New York (IEEE, New York, 1999), p. 2843.

[11] M. Reiser, *Theory and Design of Charged Particle Beams* 2nd Ed. (Wiley-VCH Inc., Weinheim Germany, 2008).

[12] L. Smith, ERDA *Summer Study for Heavy Ion Inertial Fusion*, edited by R. O. Bangerter, W. B. Herrmannsfeldt, D. L. Judd, and L. Smith, (Lawrence Berkeley Lab., Berkeley, 1976), LBL-5543, p.77-79.

[13] A. Hofmann, "Theoretical Aspects of the Behavior of Beams in Accelerators and Storage Rings", *Proceedings of the International School of Particle Accelerators*, Erice, Italy, 1976, edited by M. H. Blewett, A. Z. Zichichi, and K. Johnsen (CERN, Geneva, 1977), p. 139.

[14] A. Faltens, E.P. Lee, S.S. Rosenblum, J. of Appl. Phys., Vol. **61**, 12, p.5219, (1987).

[15] B. Beaudoin, I. Haber, R.A. Kishek, S. Bernal, T. Koeth, D. Sutter, P.G. O'Shea, and M. Reiser, Phys. Plasmas **18**, 013104 (2011).

[16] John Barnard and Steve Lund class notes on the, Interaction of Intense Charged Particle Beams with Electric and Magnetic Fields. http://hifweb.lbl.gov/NE290H/

[17] T. Koeth, B. Beaudoin, S. Bernal, I. Haber, R.A. Kishek, and P.G. O'Shea, Proc. 2011 Part Accel Conf, New York (IEEE, New York, 2011), p. 22.

[18] T.W. Koeth, B. Beaudoin, S. Bernal, I. Haber, R.A. Kishek, M. Reiser, and P.G. O'Shea, Proc. AIP Conf Proc, Annapolis (2010), p. 608.

[19] D.P. Grote, A. Friedman, I. Haber, and S. Yu, Fusion Engineering and Design, **32-33**, 193, (1996).

**Table. 1.** Beam parameters from experimental data used in both simulations and calculations.

| Peak Beam Current (mA) | Line-Charge Density (pC/m) | Initial Emittance (mm-mr) | Mean Beam radius (mm) | g-factor | $c_s$ (m/s) |
|---|---|---|---|---|---|
| 0.6 | 10.3 | 7.86 | 1.43 | 5.71 | 2.90E5 |
| 6.0 | 103.0 | 26.2 | 3.00 | 4.23 | 7.90E5 |
| 21.0 | 360.0 | 30.2 | 4.32 | 3.50 | 1.35E6 |
| 78.0 | 1300.0 | 60.5 | 7.08 | 2.51 | 2.20E6 |
| 104.0 | 1800.0 | 64.5 | 7.69 | 2.35 | 2.45E6 |